\documentclass[twocolumn,showpacs,preprintnumbers,amsmath,amssymb,floatfix,prd]{revtex4}
\usepackage{epsfig}
\begin{document}
\title{Inclusive Hadron Production in the CERN-LHC Era}
\author{Rodolfo Sassot}\email{sassot@df.uba.ar}
\affiliation{Instituto de F\'{\i}sica de Buenos Aires, CONICET, \\ Departamento de F\'{\i}sica, 
Facultad de Ciencias Exactas y Naturales, Universidad de Buenos Aires, Ciudad Universitaria, Pabell\'on\ 1 (1428) Buenos Aires, Argentina}
\author{Marco Stratmann}\email{marco@ribf.riken.jp}
\affiliation{Institut f\"{u}r Theoretische Physik,
Universit\"{a}t Regensburg, 93040 Regensburg, Germany \\
Institut f\"{u}r Theoretische Physik und Astrophysik, Universit\"{a}t W\"{u}rzburg,
97074 W\"{u}rzburg, Germany}
\author{Pia Zurita}\email{pia@df.uba.ar}
\affiliation{Instituto de F\'{\i}sica de Buenos Aires, CONICET, \\ Departamento de F\'{\i}sica, 
Facultad de Ciencias Exactas y Naturales, Universidad de Buenos Aires, Ciudad Universitaria, Pabell\'on\ 1 (1428) Buenos Aires, Argentina}

\begin{abstract}
We present a detailed phenomenological analysis of single-inclusive hadron production 
at the CERN-LHC in both proton-proton and proton-lead collisions.
First data from the LHC experiments on charged hadron spectra are
compared to next-to-leading order QCD expectations, and predictions
are made for identified pion, kaon, and proton distributions differential
in transverse momentum and rapidity for LHC energies from 900 GeV to 14 TeV.
The results are obtained with the latest sets of vacuum fragmentation functions based on
global QCD analyses, and recently proposed medium modified fragmentation functions 
are used to model hadronization in proton-lead collisions assuming standard QCD factorization.
Besides estimating theoretical ambiguities due to the choice of factorization and renormalization
scales and parton densities, we carefully assess uncertainties due to our present knowledge 
of parton-to-hadron fragmentation functions with the Lagrange multiplier technique.  
It is outlined to what extent future LHC data will contribute to further our 
quantitative understanding of hadronization processes.
\end{abstract}


\maketitle

%%%%%%%%%%%%%%%%%%%%%%%%%%%%%%%%%%%%%
\section{Introduction and Motivation}
%%%%%%%%%%%%%%%%%%%%%%%%%%%%%%%%%%%%%
Single-inclusive hadron production is increasingly becoming a precise and versatile 
tool to quantitatively study various aspects of Quantum Chromodynamics (QCD), 
supplementing the insights already gained from fully inclusive observables.
First and foremost, precise data obtained in electron-positron, lepton-nucleon, and
hadron-hadron collisions at different center-of-mass system (c.m.s.) energies $\sqrt{S}$
have established collinear factorization \cite{ref:fact} as the foundation
for a perturbative QCD (pQCD) approach to hard scattering processes 
involving identified hadrons produced at large transverse momentum $p_T$.

Within the framework of collinear factorization, the presence of a hard scale $p_T$ allows 
one to compute with remarkable phenomenological success various kinematic distributions 
of the observed final-state hadron $H$ as a convolution of partonic short-distance 
scattering cross sections, calculable as a perturbative 
expansion in the strong coupling $\alpha_s$, and universal but non-perturbative 
functions capturing the long-distance physics as represented by parton distribution and 
fragmentation functions.

By now, information on these non-perturbative inputs is routinely extracted in 
``global QCD analyses" by consistently comparing data sets from many different experiments 
and processes with theoretical expectations at a given order in perturbation theory.
In particular for parton density functions (PDFs), such type of analyses have reached 
a high level of sophistication in recent years \cite{ref:cteq,ref:mstw,ref:nnpdf}. Various
methods have been put forward to arrive at faithful estimates of the remaining uncertainties 
in our understanding of the hadronic structure \cite{ref:hessian,ref:lagrange,ref:nnpdf}
and how they affect, e.g., the level at which we have control of standard model background 
processes at the CERN Large Hadron Collider (LHC).

Extractions of parton-to-hadron fragmentation functions (FFs) and their uncertainties are 
a much more recent achievement \cite{ref:dsspion,ref:dssproton,ref:akk}, mainly hampered 
by the considerably smaller amount of experimental results available. In general, reducing 
uncertainties of FFs is of utmost importance to further our understanding of the hadronization 
process, test the scale dependence of FFs as predicted in pQCD, and to delineate kinematic 
regions where factorized, color-independent, collinear parton-to-hadron fragmentation 
provides a sensible approximation.

The limited knowledge of FFs has a serious impact on different studies, 
such as, for instance, global QCD analyses of data taken in collisions of longitudinally 
polarized nucleons and leptons aiming to address the fundamental question of how the 
spin of the nucleon is composed of the intrinsic spins and orbital angular momenta of 
quarks and gluons \cite{ref:dssv}. Since many of the data available so far involve identified 
hadrons in the final-state, either in semi-inclusive deep-inelastic lepton-nucleon or in 
proton-proton scattering processes, FFs are a crucial ingredient of theoretical 
analyses of the spin structure of the nucleon. 
Processes with identified hadrons are also increasingly 
used as decisive probes to gain insight about the properties 
and nature of cold and hot nuclear matter in heavy ion collisions both 
at the BNL Relativistic Heavy Ion Collider (RHIC) and soon also at the LHC. 
Again, precise knowledge of FFs and possible modifications induced by the nuclear
medium are crucial ingredients that have received increasing attention recently
\cite{ref:nuclreview}.

In the following, we will assess in some detail how upcoming data from the LHC
on high $p_T$ hadron production in an unprecedented kinematic regime
will help to further our knowledge of FFs both in the vacuum and in a nuclear environment.
To this end, we first need to discuss briefly some theoretical preliminaries such as
the formalism for single-inclusive hadron production within pQCD factorization, the range of
applicability of universal vacuum FFs, estimates and propagation of uncertainties,
and a recent proposal for medium modified FFs.

%%%%%%%%%%%%%%%%%%%%%%%%%%%%%%%%%%%%%%%%%%%%%%%%%%%%%%
\subsection{Framework for hadron production at the LHC}
%%%%%%%%%%%%%%%%%%%%%%%%%%%%%%%%%%%%%%%%%%%%%%%%%%%%%%
%
Throughout this paper, we are interested in the single-inclusive invariant cross section 
for the production of a hadron $H$ with energy $E$ and momentum $\vec{p}$
in hadron-hadron or hadron-nucleus collisions. 
Assuming pQCD factorization, the relevant theoretical expression 
for $pp$ collisions at the LHC schematically reads
\begin{eqnarray}
\nonumber
E\frac{d^3\sigma^H}{d\vec{p}} &=&
\sum_{a,b,c} f_a(x_a,\mu_f) \otimes f_b(x_b,\mu_f)
\otimes D_c^H(z_c,\mu_{f^{\prime}})  \\
&&
\otimes \;d\hat{\sigma}_{ab\to cX}(S,\alpha_s,x_a,x_b,z_c,\mu_f,\mu_r,\mu_{f^\prime}),
\label{eq:xsec}
\end{eqnarray}
where the sum is over all contributing partonic subprocesses $ab\rightarrow cX$
contained in the perturbatively calculable
short-distance scattering cross sections $d\hat{\sigma}_{ab\to cX}$. 

The scales $\mu_f$ and $\mu_{f^{\prime}}$ are introduced to
factorize initial and final-state collinear singularities into the 
scale dependent PDFs and FFs, $f_{a,b}(x_{a,b},\mu_f)$ and $D_c^H(z_c,\mu_{f^{\prime}})$,
respectively.
$\mu_r$ denotes the energy scale at which $\alpha_s$ is being renormalized.
The residual dependence of Eq.~(\ref{eq:xsec}) on the arbitrary scales 
$\mu_{f,{f^{\prime}},r}$ can be taken as a measure for the theoretical ambiguity due
to the truncation of the perturbative series at a given fixed order in $\alpha_s$. 
To estimate its impact, we will follow the usual procedure and vary the scales within a
factor of two around the default choice $p_T$. It turns out that scale
variations are the dominant theoretical uncertainty for hadron production 
at the LHC.

$x_{a,b}$ are the fractions of longitudinal momentum of the colliding hadrons
taken by the interacting partons $a$ and $b$. 
Similarly, $z\equiv z_c$ denotes the collinear momentum fraction of the fragmenting parton $c$
carried by the produced hadron $H$. 
Neither $x_{a,b}$ nor $z$ are measurable quantities, and any given data point 
characterized by the $p_T$ of the hadron $H$ and the c.m.s.\ energy $\sqrt{S}$ probes
both the PDFs and the FFs at a different range of momentum fractions and 
scales of ${\cal{O}}(p_T)$. 
Likewise, different hadron species $H=\pi^0,\pi^{\pm},K^{\pm},\ldots$ determine
complementary aspects of the hadronization process, i.e., different FFs $D_c^H$.
The relevance of each parton flavor $c$ depends on the quark content of $H$ and
on $p_T$ and $\sqrt{S}$, which control the contributions of the various partonic
channels $ab\to cX$ to the sum in Eq.~(\ref{eq:xsec}).
We will demonstrate how the different partonic subprocesses are expected to contribute
to hadron production yields at LHC energies.

A noteworthy property of the invariant cross section in Eq.~(\ref{eq:xsec})
is its approximate power-law behavior \cite{ref:xt,ref:arleo}
\begin{equation}
\label{eq:xtscaling}
\sigma_{\mathrm{inv}}\equiv E\frac{d^3\sigma^H}{d\vec{p}} = F(x_T)/p_T^{n(x_T,\sqrt{S})}
\end{equation}
for fixed $x_T\equiv 2p_T/\sqrt{S}$.
In the naive, scale-invariant parton model one expects scaling with $n=4$.
The running of $\alpha_s$ and the scaling violations of the
PDFs and FFs as predicted by pQCD lead to deviations from exact scaling
for Eqs.~(\ref{eq:xsec}) and (\ref{eq:xtscaling}), i.e., $n=n(x_T,\sqrt{S})$,
which can be explored by comparing, e.g., $x_T$ hadron spectra at different
$\sqrt{S}$. We shall briefly touch upon $x_T$ scaling at LHC energies and
estimate theoretical scale uncertainties in predicting $n=n(x_T,\sqrt{S})$.

%%%%%%%%%%%%%%%%%%%%%%%%%%%%%%%%%%%%%%%%%%%%%%%%%%%%
\subsection{Applicability of FFs and uncertainty estimates}
%%%%%%%%%%%%%%%%%%%%%%%%%%%%%%%%%%%%%%%%%%%%%%%%%%%%
%
Since the LHC sets a new energy frontier for hadron production,
it is crucial to first convince ourselves that the accessible range of $z$ 
is still compatible with the applicability of the concept of FFs
\cite{ref:ffdef} within the factorized framework as outlined above. 
Contrary to PDFs, where the small $x_{a,b}$ regime is rather well explored
down to momentum fractions of ${\cal{O}}(10^{-4})$ \cite{ref:cteq,ref:mstw,ref:nnpdf}
and has been subjected to a very detailed theoretical scrutiny,
the phenomenological access to FFs is much more restricted.
Neglected hadron mass effects, potential higher twist corrections, and instabilities 
in the timelike scale evolution limit the usage of FFs to rather large values of 
$z\gtrsim 0.05$; see, e.g.\ Ref.~\cite{ref:dsspion}. 
We shall show that for all practical applications at the LHC the condition $z\gtrsim 0.05$ 
is well met as the bulk of the cross section for inclusive hadron production 
samples on average large momentum fractions, $\langle z\rangle \simeq 0.5$, well above the
kinematic lower limit $z \approx 2p_T/\sqrt{S}$.

Upcoming experimental results for single-inclusive hadron production at the LHC 
can be straightforwardly included \cite{ref:mellin} in existing global QCD analyses of FFs 
\cite{ref:dsspion,ref:dssproton}, which are the most efficient and consistent method 
to deconvolute information on these non-perturbative functions from the interplay
of various observables measured at different energy scales.  
The large range of transverse momenta $p_T$ accessible at the LHC 
will allow for unprecedented studies of evolution effects for FFs.
We note that for both the timelike scale evolution of FFs \cite{ref:evol} and the partonic
hard scattering cross sections $d\hat{\sigma}_{ab\to cX}$ \cite{ref:xsecnlo} in Eq.~(\ref{eq:xsec}),
pQCD calculations at next-to-leading order (NLO)
accuracy are ``state-of-the-art" and used throughout this work.
In any case, they are mandatory for an
accurate and meaningful comparison of theory and data
due to often sizable NLO QCD corrections and, in particular, to allow for estimates
of scale uncertainties.

As mentioned above, besides establishing a small set of ``best-fit" 
parameters in a $\chi^2$ minimization to model the functional form
of the FFs for different flavors and hadrons, assessing their uncertainties and propagating them 
to physical observables is an equally important goal. 
Here, the most robust technique is based on 
Lagrange multipliers \cite{ref:lagrange,ref:dsspion,ref:dssproton}
which makes no assumptions about the behavior of the $\chi^2$ profile near its minimum.
We illustrate the usefulness of this method by estimating the uncertainties 
from FFs for charged hadron production at the LHC and 
Fermilab's Tevatron $p\bar{p}$ collider.
For the latter, data from the CDF collaboration \cite{ref:cdfdata} 
have recently caused some stir as the measured cross section
above $p_T\gtrsim 20\,\mathrm{GeV}$ exceeds theoretical expectations 
by orders of magnitude \cite{ref:cdffuzz}. Since the data are also in excess of 
single-inclusive jet cross section measurements, the most likely explanation 
seems to be some experimental problem \cite{ref:cdffuzz}.
Nevertheless, we believe it is an useful exercise to evaluate 
the theoretical ambiguities caused by FFs also in this case. 
It turns out that naive estimates, for instance, by comparing the results obtained
with two different sets of FFs, often seriously underestimate uncertainties from FFs. 

%%%%%%%%%%%%%%%%%%%%%%%%%%%%%%%%%%%%%%%%%%%%%%%%%%%%%%%%%
\subsection{FFs in a nuclear environment at the LHC}
%%%%%%%%%%%%%%%%%%%%%%%%%%%%%%%%%%%%%%%%%%%%%%%%%%%%%%%%%
%
It is well known that results for hadron production processes
occurring in a nuclear medium can differ significantly from 
similar experiments involving only light nuclei or proton targets,
showing both suppression and enhancement of the rates 
depending on the details of the observable.
Recent examples include production rates of pions and kaons 
in semi-inclusive deep-inelastic scattering off different nuclei
as provided by the HERMES experiment \cite{ref:hermes}
and in deuteron-gold collisions measured at RHIC \cite{ref:rhic-dau}.
The origin of the observed nuclear modifications
has been attributed to a variety of conceivable mechanisms and models 
\cite{ref:nuclreview}
besides the well-known modification of parton densities in nuclei (nPDFs)
\cite{ref:nds,ref:hirai,ref:eps}.
Available models incorporate ideas based on interactions 
between the nuclear medium and, e.g., the final-state hadron
or the seed partons before the hadronization takes place,
and reproduce, with different degree of success, some features of the data;
for recent reviews, see Ref.~\cite{ref:nuclreview}.

nPDFs provide an effective and phenomenologically successful
way to factorize the influence of the nuclear environment 
on the interacting partons into sets of universal functions 
which scale in energy like ordinary PDFs and can be obtained in
global QCD fits to available data \cite{ref:nds,ref:hirai,ref:eps}.
The quite natural extension of this idea to final-state nuclear effects
has been put forward only very recently by introducing the concept
of medium modified fragmentation functions (nFFs) \cite{ref:nuclreview}.
First QCD fits for identified pions and kaons were 
provided in Ref.~\cite{ref:nff} recently. 
As for nPDFs, it was demonstrated that within the precision 
of the available data, universal nFFs are a viable concept and
factorization similar to Eq.~(\ref{eq:xsec}) holds at least approximately
\cite{ref:nff} despite being much more speculative than in $pp$ collisions
\cite{ref:breaking}.
The nuclear $A$ dependence of the nFFs can be most economically
parametrized in a convolution approach \cite{ref:nff}
\begin{equation}
D_{c/A}^H(z,\mu_0) =
\int_z^1 \frac{dy}{y} W_c^H(y,A,\mu_0)D_c^H\left(\frac{z}{y},\mu_0\right)
\label{eq:conv-ansatz}
\end{equation}
which relates the nFFs $D_{c/A}^H$ to the fairly well known vacuum FFs
of DSS \cite{ref:dsspion} at some initial scale $\mu_0$ 
through a weight function $W_c^H$
with only a small amount of extra parameters. nFFs at scales
$\mu>\mu_0$ are then obtained by applying the 
standard timelike evolution equations \cite{ref:evol}.

Combined with nPDFs, nFFs allow one to treat a large class
of hard hadron production processes where a nucleus collides
with a lepton or a nucleon (light nucleus) in a consistent pQCD
framework based on factorization. 
Exploiting the predictive power of the factorized approach, we will provide
predictions for pion production in future proton-lead ($pPb$) collisions at the LHC
in a wide range of $p_T$ and rapidity.
$pPb$ collisions at the LHC are conceivable 
up to a c.m.s.\ energy of $\sqrt{S}\approx 8.8\,\mathrm{TeV}$ \cite{ref:lhcppb}
though not part of the initial LHC physics program.
In addition, we shall discuss how the admixture of the different
contributing partonic subprocesses $ab\to cX$ and the range of probed momentum fractions
$z$ is expected to change in a nuclear environment at LHC energies.
Such measurements will be crucial to further our knowledge
of hadronization in a nuclear medium by exploring to what extent
factorization breaking effects due to interactions
of partons with the medium come into play and limit the usefulness
of nPDFs and nFFs. Comparisons of hadron rates obtained $pp$ and $pPb$ collisions
will also help to unravel and understand
the properties of hot and dense QCD matter.

The remainder of the paper is organized as follows:
in the next Section, we compare first results for $p_T$ spectra
of unidentified charged hadrons from ATLAS \cite{ref:atlas} and
CMS \cite{ref:cms} with pQCD calculations at NLO accuracy. 
In Sec.~III, we present expectations for both identified and
unidentified hadron production cross sections in $pp$ collisions at the LHC 
in a broad range of $p_T$ and rapidity $y$.
The results are supplemented by studies of theoretical scale and PDF 
ambiguities, the relevance of different partonic subprocesses, and the ranges of
momentum fractions $x_{a,b}$ and $z$ predominantly probed. 
We touch upon $x_T$ scaling and give estimates of FF uncertainties for
charged particle yields at the LHC and the Tevatron using the Lagrange multiplier technique. 
Pion production in $pPb$ collisions and medium modified FFs are discussed 
in Sec.~IV. We summarize our main results in Sec.~V.

%%%%%%%%%%%%%%%%%%%%%%%%%%%%%%%%%%%%%%%%%%%%%%%%%%%%%%%%%%
\section{\label{sec:firstdata} Comparison to first LHC data}
%%%%%%%%%%%%%%%%%%%%%%%%%%%%%%%%%%%%%%%%%%%%%%%%%%%%%%%%%%
%
Before turning to a detailed discussion of theoretical
expectations for single-inclusive hadron production at LHC
energies in Sec.~\ref{sec:pp}, we take a brief look at
first results from ATLAS \cite{ref:atlas} and
CMS \cite{ref:cms} for $p_T$ differential charged hadron yields. 

Charged hadron multiplicities are the first results 
of the LHC physics program, and data were reported shortly 
after the startup of the LHC by the ALICE \cite{ref:alicemult,ref:alicept}, 
ATLAS \cite{ref:atlas}, and CMS \cite{ref:cms}
experiments which impressively demonstrated the readiness
of their detector systems.
They provide measurements of the number of charged hadrons
$N_{ch}$ with respect to (w.r.t.) their transverse momentum
$p_T$ and pseudorapidity $\eta$.
The main result so far is the increase of the
pseudorapidity density $dN_{ch}/d\eta$
at central rapidities with the 
c.m.s.\ energy of the $pp$ collisions in the range
$\sqrt{S}=0.9 - 7\,\mathrm{TeV}$ \cite{ref:alicemult,ref:cms}.

The pseudorapidity density
$dN_{ch}/d\eta$ is not amenable to pQCD calculations
based on factorization, Eq.~(\ref{eq:xsec}), as the bulk of the 
produced hadrons has very low $p_T$, well below $1\,\mathrm{GeV}$,
and a hard scale is lacking.
However, both ATLAS \cite{ref:atlas} and CMS \cite{ref:cms} also present data for 
$dN_{ch}/dp_T$ in the rapidity range $|\eta|\le 2.5$ and
$|\eta|\le 2.4$, respectively, at sufficiently high $p_T$.
Very recently, also ALICE published a $p_T$ spectrum at 
$\sqrt{S}=900\,\mathrm{GeV}$ and $|\eta|\le 0.8$ \cite{ref:alicept},
but data tables are not yet available for comparison to theory.
In each case, the charged particle multiplicities are normalized
to the number of inelastic non-single-diffractive (NSD) interactions
but based on slightly different event selections.
%
%%%%%%%%%%%%%%%%%%%%%%%%%%%%%%%
% FIGURE 1: ATLAS 900GeV
%%%%%%%%%%%%%%%%%%%%%%%%%%%%%%%
%
\begin{figure}[th]
\begin{center}
\vspace*{-0.6cm}
\epsfig{figure=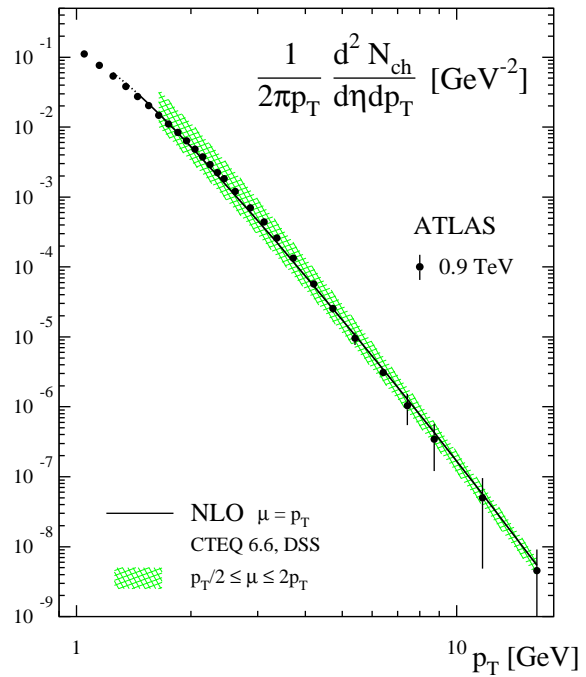,width=0.48\textwidth}
\end{center}
\vspace*{-0.5cm}
\caption{\label{fig:atlas-data}
Comparison of the single-inclusive charged hadron yield per NSD event in 
the rapidity range $|\eta|\le 2.5$ in $pp$ collisions
at $\sqrt{S}=900\,\mathrm{GeV}$ measured by ATLAS \cite{ref:atlas}
with a NLO pQCD calculation using the DSS FFs \cite{ref:dssproton}
and CTEQ6.6 PDFs \cite{ref:cteq} (solid line). 
The shaded band indicates the theoretical uncertainty from varying
the scales in Eq.~(\ref{eq:xsec}) by a factor of 2 
around the default choice $\mu_f=\mu_{f^{\prime}}=\mu_r=p_T$.
Note that the normalization of the curve is determined by a fit;
see text.}
\end{figure}

\begin{figure}[th]
\begin{center}
\vspace*{-0.6cm}
\epsfig{figure=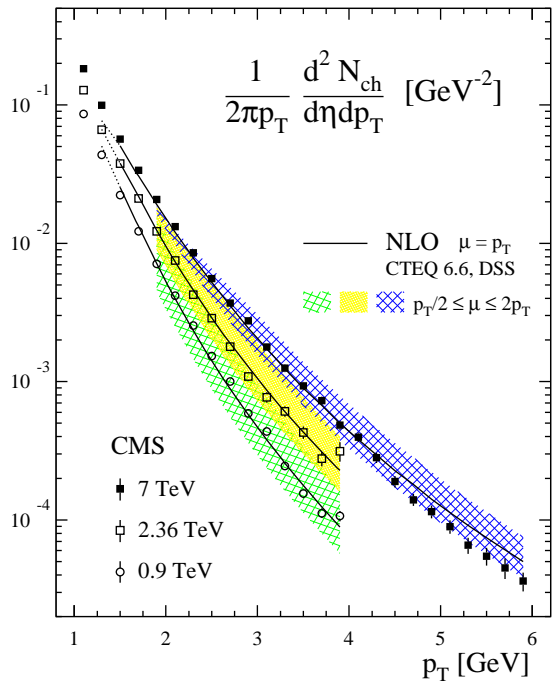,width=0.48\textwidth}
\end{center}
\vspace*{-0.5cm}
\caption{As in Fig.~\ref{fig:atlas-data} but
now in the rapidity range $|\eta|\le 2.4$ 
and for three different c.m.s.\ energies 
$\sqrt{S}=0.9$, 2.36, and $7\,\mathrm{TeV}$
as measured by the CMS experiment \cite{ref:cms}. 
\label{fig:cms-data}}
\end{figure}

In order to compare the obtained invariant yields $1/(2\pi p_T) d^2N_{ch}/d\eta dp_T$
with pQCD predictions for $E d^3\sigma^H/d\vec{p}$ in Eq.~(\ref{eq:xsec}), 
one needs to normalize them with the total NSD cross section $\sigma_{NSD}$,
which unfortunately is not specified in Refs.~\cite{ref:atlas,ref:cms,ref:alicept}. However, we
estimate $\sigma_{NSD}$ for each available data set by scaling the theoretical
single-inclusive hadron yields at NLO by $1/\sigma_{NSD}$ and fitting a common, 
i.e., $p_T$-independent, value for each experiment. 
We use the CTEQ6.6 set of PDFs \cite{ref:cteq} and the DSS parton-to-unidentified 
charged hadron FFs \cite{ref:dssproton} in Eq.~(\ref{eq:xsec}) 
and choose $\mu_f=\mu_{f^{\prime}}=\mu_r=\mu=p_T$.
The results of the NLO calculations are shown as solid lines 
in Figs.~\ref{fig:atlas-data} and \ref{fig:cms-data} 
and compared to ATLAS \cite{ref:atlas} and CMS \cite{ref:cms} data, respectively.
Note that only data with $p_T\ge 1\,\mathrm{GeV}$ are displayed. 
We refrain from giving theoretical expectations 
based on Eq.~(\ref{eq:xsec}) for $p_T\lesssim 1.5\,\mathrm{GeV}$ 
where power suppressed corrections to the factorized 
pQCD framework and other non-perturbative soft contributions are relevant.
In any case, all sets of PDFs \cite{ref:cteq} and FFs \cite{ref:dssproton} 
are not applicable for too small scales $\mu\simeq p_T$.

The shaded bands in Figs.~\ref{fig:atlas-data} and \ref{fig:cms-data}
give an indication of the theoretical uncertainties due to the 
truncation of the perturbative series in Eq.~(\ref{eq:xsec}) 
at NLO accuracy. As is customary, they are obtained by 
simultaneously varying all scales in Eq.~(\ref{eq:xsec}) by a factor of 2 
around the default choice $\mu_r=p_T$. 
Given the fact that the normalization of the theoretical results has to
be determined by a fit, we refrain from studying other, usually subleading,
sources of uncertainties like variations of PDF sets, at this point. 
%
%%%%%%%%%%%%%%%%%%%%%%%%%%%%%%%%%%
% FIGURE 2: CMS 0.9, 2.36, 7 TeV
%%%%%%%%%%%%%%%%%%%%%%%%%%%%%%%%%%
%

The overall agreement between the $p_T$ dependence as predicted by pQCD calculations at 
NLO accuracy and data
is excellent, and the drop of the cross sections with increasing $p_T$
over many orders of magnitude is accurately reproduced. 
As was already noticed in $pp$ collisions at BNL-RHIC at lower c.m.s.~energies,
NLO pQCD calculations at the leading-twist level tend to describe 
single inclusive hadron yields well down to fairly low values of 
$p_T\simeq 1.5\,\mathrm{GeV}$. 
We note a mild tension with CMS data \cite{ref:cms} in Fig.~\ref{fig:cms-data}
for $\sqrt{S}=7\,\mathrm{TeV}$ and $p_T\gtrsim 4.5\,\mathrm{GeV}$, which are 
all at the lower edge of the scale uncertainty band.
Notice that of all hadron production data, the ones taken 
at $\sqrt{S}=7\,\mathrm{TeV}$ probe the smallest $x_{a,b}$ values 
in the PDFs so far, down to a kinematic limit of about $x_T\simeq 5\cdot 10^{-4}$.

Upcoming precision data from the LHC will decisively test all theoretical 
aspects of inclusive hadron production in an 
unprecedented energy range, from non-perturbative
PDFs and FFs to the validity of the leading-twist approximation in 
Eq.~(\ref{eq:xsec}).
In addition to unidentified charged hadron yields measured so far, 
the ALICE experiment is designed to identify various hadron species
like pions, kaons, and protons at central rapidities. 
These results will provide a vital input for future global analyses of FFs.
Detailed theoretical expectations for single inclusive 
hadron production at the LHC will be discussed in the
next Section along with estimates of uncertainties.

%%%%%%%%%%%%%%%%%%%%%%%%%%%%%%%%%%%%%%%%%%%%%%%%%%%%%%%%%%%%%%
\section{\label{sec:pp}Expectations for single-inclusive
high-$p_T$ hadron production in $pp$ collisions at ${\mathrm{TeV}}$ energies}
%%%%%%%%%%%%%%%%%%%%%%%%%%%%%%%%%%%%%%%%%%%%%%%%%%%%%%%%%%%%%%
%
%%%%%%%%%%%%%%%%%%
% FIGURE 3: PI+ ENERGY DEPENDENCE AND PI+/PI- RATIO
%%%%%%%%%%%%%%%%%%
\begin{figure}[th]
\begin{center}
\vspace*{-0.6cm}
\epsfig{figure=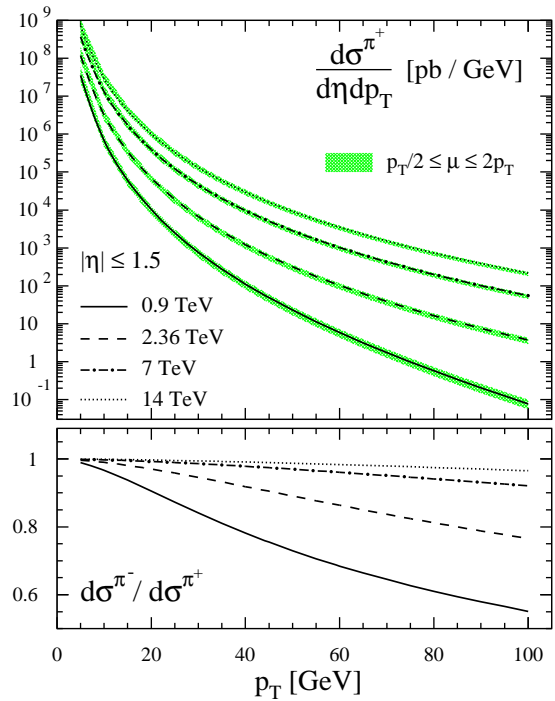,width=0.48\textwidth}
\end{center}
\vspace*{-0.5cm}
\caption{Upper panel: NLO cross section for $\pi^+$ production at
four different c.m.s.\ energies, integrated in the pseudorapidity range
$|\eta|\le 1.5$. The shaded bands indicate the uncertainty 
associated with the variation of the scales in the range
$p_T/2 \le \mu_{f,{f^{\prime}},r} \le 2p_T$.
Lower panel: ratio of the $\pi^-$ and $\pi^+$ yields.
\label{fig:piplus}}
\end{figure}
Encouraged by the first results from the LHC experiments and the successful
comparison to theoretical expectations based on the DSS set of fragmentation
functions \cite{ref:dsspion,ref:dssproton}, 
we now turn to a detailed discussion of both identified and unidentified 
hadron production cross sections to be studied at the LHC in the future. 
While ATLAS, CMS, and ALICE can all measure unidentified charged
hadron yields, only ALICE is capable of tagging different hadron species at
central rapidities, which is crucial for detailed tests of our
current knowledge of FFs and the underlying framework of pQCD.

Our results are supplemented by estimates of theoretical scale, PDF, and FF
uncertainties. To elucidate the impact of future LHC data on our understanding
of pQCD factorization and, in particular, hadronization,
we discuss the relevance of different partonic subprocesses
$d\sigma_{ab\rightarrow cX}$ in Eq.~(\ref{eq:xsec}), the role of quark and gluon
fragmentation, and the ranges of momentum fractions 
which will be predominantly probed. 

If not stated otherwise, we use the CTEQ6.6 set of PDFs
and the associated Hessian sets for PDF uncertainty estimates \cite{ref:cteq}.
The parton-to-hadron FFs are taken from DSS \cite{ref:dsspion,ref:dssproton},
and uncertainties related to FFs will be estimated with the Lagrange multiplier method
\cite{ref:lagrange,ref:dsspion,ref:dssproton}.
All calculations are performed at NLO accuracy with hard
scattering cross sections, PDFs, and FFs in Eq.~({\ref{eq:xsec})
evaluated in the $\overline{\mathrm{MS}}$ scheme.
We take the transverse momentum $p_T$ of the produced hadron as
the default choice for the factorization and renormalization scales,
$\mu_{f,{f^\prime}}$ and $\mu_r$, respectively, in Eq.~({\ref{eq:xsec}).
To estimate theoretical ambiguities associated with the truncation of
the perturbative series at NLO accuracy we vary all scales by a factor
of two up and down their central value $p_T$ as is commonly done. 

%%%%%%%%%%%%%%%%%%%%%%%%%%%%%%%%%%%%%%%%%%%%%%%%%%%%%%%%%%%%%%%%%%%%%%%%%%
\subsection{Pion, kaon, and proton production at the LHC}
%%%%%%%%%%%%%%%%%%%%%%%%%%%%%%%%%%%%%%%%%%%%%%%%%%%%%%%%%%%%%%%%%%%%%%%%%%
%
%%%%%%%%%%%%%%%%%%%%%%%%%%%%%%%%%%%%%%%%%%%%%%%%%%%%%%%%%%%%%%%%
% FIGURE 4: pi+, K+, proton yields at 7TeV & PDF uncertainties
%%%%%%%%%%%%%%%%%%%%%%%%%%%%%%%%%%%%%%%%%%%%%%%%%%%%%%%%%%%%%%%
\begin{figure}[th]
\begin{center}
\vspace*{-0.6cm}
\epsfig{figure=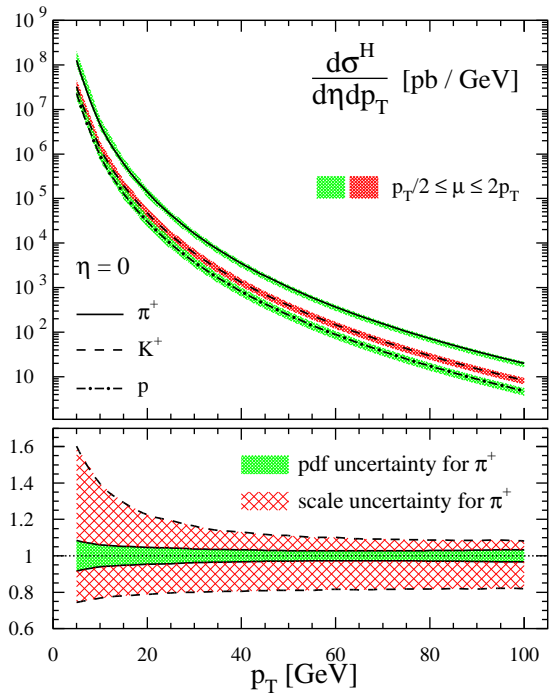,width=0.48\textwidth}
\end{center}
\vspace*{-0.5cm}
\caption{Upper panel: comparison of the yields for $\pi^+$, $K^+$,
and $p$ at NLO accuracy for pseudorapidity $\eta=0$ and $\sqrt{S}=7\,\mathrm{TeV}$.
As before, the shaded bands give an estimate of the theoretical scale ambiguity.
Lower panel: PDF uncertainty for $\pi^+$ production 
w.r.t.\ the solid line in the upper panel, estimated with the
Hessian eigenvector sets for CTEQ6.6 \cite{ref:cteq} (solid band) and compared
to the ambiguity from scale variations (hatched band).
\label{fig:plus}}
\end{figure}
We start off with showing the energy dependence of the $p_T$ differential 
cross section for charged pion production integrated in the pseudorapidity 
range $|\eta|\le 1.5$ in the upper panel of Fig.~\ref{fig:piplus} 
for $\sqrt{S}= 0.9,\,2.36,\, 7,$ and $14\,\mathrm{TeV}$. 
The shaded bands give an estimate of the theoretical ambiguities due
the variations of $\mu_{f,f^\prime,r}$ as described above.
The typical scale uncertainty for $\sqrt{S}= 14\,\mathrm{TeV}$
amounts to about $25\%$ at $p_T\approx 20\,\mathrm{GeV}$
and $14\%$ at $p_T\approx 100\,\mathrm{GeV}$. At the same values of $p_T$
but lower c.m.s.~energies, $\sqrt{S}= 0.9\,\mathrm{TeV}$, the corresponding
uncertainties can reach up to $30\%$.  

The lower panel shows the expected difference of $\pi^+$ and $\pi^-$ 
yields as predicted by the DSS set of FFs \cite{ref:dsspion}.
We recall that at present, the charge separation of FFs is mainly constrained
by semi-inclusive deep-inelastic scattering (SIDIS) data at a 
relatively low scale $Q^2\simeq 2.5\mathrm{GeV}$
and to a much lesser extent by data from RHIC \cite{ref:dsspion}.
As a consequence, uncertainties are still sizable in current sets of FFs.
As can be seen from Fig.~\ref{fig:piplus},
the ratio $d\sigma^{\pi^{-}}/d\sigma^{\pi^{+}}$ drops only very slowly
with increasing $p_T$ for both $\sqrt{S}= 14$ and $7\,\mathrm{TeV}$.
This is due to the dominance of gluon initiated
hard scattering and hadronization processes, which prevails up to fairly 
high $p_T$; see Figs.~\ref{fig:pp-support}~(a) and (c) and the discussions below.
Clearly, a precise measurement of $d\sigma^{\pi^{-}}/d\sigma^{\pi^{+}}$ is of
great phenomenological importance 
but at the same time also very challenging at nominal LHC energy as sufficiently large integrated 
luminosities are required to resolve effects 
of a few percent at high $p_T$.
Lower c.m.s.\ energies, $\sqrt{S}=2.36$ and $0.9\,\mathrm{TeV}$ are more favorable
as they probe larger momentum fractions $x_{a,b}$ in the nucleon 
in the same range of $p_T$, i.e., larger $x_T\equiv2p_T/\sqrt{S}$, 
where quark initiated scattering processes more and more dominate.

The expectation that $d\sigma^{\pi^{+}}>d\sigma^{\pi^{-}}$ can be understood 
by inspecting the role of different partonic subprocesses in $pp$ collisions;
see Fig.~\ref{fig:pp-support}~(a) below.
Quark-gluon scattering is the second most important channel for hadron production.
The abundance of $u$ quarks in a proton and the $u\bar{d}$ valence flavor 
structure of a $\pi^+$, i.e., $D_u^{\pi^{+}}>D_{u}^{\pi^{-}}$, explains
the observed hierarchy for the $\pi^{\pm}$ production yields found
in Fig.~\ref{fig:piplus}. 

Figure~\ref{fig:plus} compares the cross sections for $\pi^+$, $K^+$, and
proton production as a function of $p_T$ for $\sqrt{S}=7\,\mathrm{TeV}$
and $\eta=0$.
At $p_T\approx20\,\mathrm{GeV}$ pions yields are roughly a factor of
3 larger than those for $K^+$, and the production of protons is
suppressed even further.
Together, pions, kaons, and protons account
for almost the entire yield of charged hadrons; see
Fig.~\ref{fig:lhc-hadron}.
The mixture of hadron species is largely independent of
$p_T$, with the fraction of pions decreasing slightly as $p_T$ increases.
This is because the mean value of $z$ changes only very slowly with
$p_T$, see Fig.~\ref{fig:pp-support}~(e). At higher $z$, fragmentation into
heavier hadrons like kaons and protons is somewhat 
enhanced w.r.t.~those into pions \cite{ref:dsspion,ref:dssproton}.

The typical PDF uncertainty for hadron production at the LHC is illustrated in
the lower panel of Fig.~\ref{fig:plus} and amounts to about $5\%$ at $p_T=20\,\mathrm{GeV}$.
It is computed with the help of the 44 Hessian eigenvector sets provided by
CTEQ \cite{ref:cteq} and compared to variations of the cross section
due to the choice of scales $\mu_{f,{f^{\prime}},r}$ in Eq.~(\ref{eq:xsec}).
Because of the dominance of gluon initiated scattering processes,
see Fig.~\ref{fig:pp-support}~(a), the PDF uncertainty
reflects to a large extent the present ambiguity in the gluon PDF at the relevant scale
$\mu\simeq p_T$ and range of momentum fraction $x$.
In particular at smaller values of $p_T$, the theoretical ambiguity due
to the truncation of the pQCD series at NLO turns out to be
by far the most relevant one.
 
%%%%%%%%%%%%%%%%%%%%%%%%%%%%%%%%%%%%%%%%%%%%%%%%%%%%%%%%%%%%%%%%
% FIGURE 5: pi0 rapidity dep. at 7TeV 
%%%%%%%%%%%%%%%%%%%%%%%%%%%%%%%%%%%%%%%%%%%%%%%%%%%%%%%%%%%%%%%
\begin{figure}[th]
\begin{center}
\vspace*{-0.6cm}
\epsfig{figure=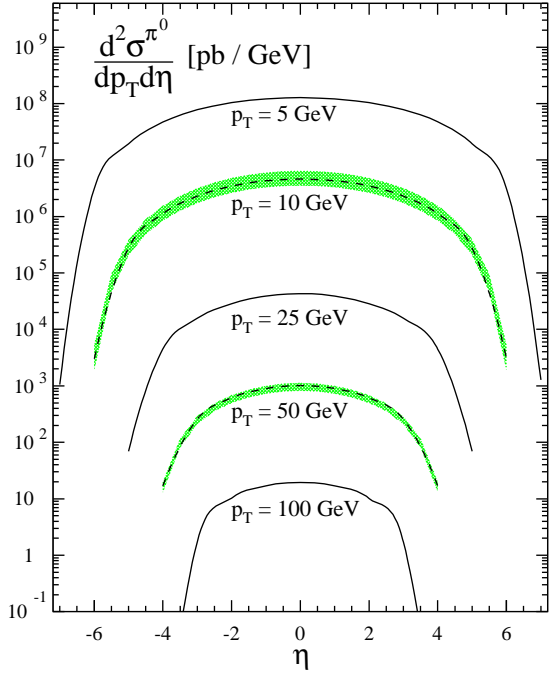,width=0.48\textwidth}
\end{center}
\vspace*{-0.5cm}
\caption{Pseudorapidity dependence of the neutral pion production cross section
at $\sqrt{S}=7\,\mathrm{TeV}$ for various fixed values of $p_T$.
The typical theoretical scale uncertainty is indicated by the shaded bands
for $p_T=10$ and $50\,\mathrm{GeV}$.
\label{fig:pi0rap}}
\end{figure}
The pseudorapidity dependence of the production cross section for neutral pions 
at $\sqrt{S}=7\,\mathrm{TeV}$ is shown in Fig.~\ref{fig:pi0rap} for various fixed 
values of $p_T$. Again, the typical scale uncertainty is indicated by the 
shaded bands for $p_T=10$ and $50\,\mathrm{GeV}$.
Measurements of the $\eta$ dependence for fixed $p_T$ in a wide range are of great 
phenomenological interest 
as they emphasize different partonic subprocesses and momentum fractions
$x_{a,b}$ and $z$ as compared to the $p_T$ differential yields shown in
Figs.~\ref{fig:piplus} and \ref{fig:plus}.

%%%%%%%%%%%%%%%%%%%%%%%%%%%%%%%%%%%%%%%%%%%%%%%%%%%%%%%%%%%%%%%%
% FIGURE 6: support plot 
%%%%%%%%%%%%%%%%%%%%%%%%%%%%%%%%%%%%%%%%%%%%%%%%%%%%%%%%%%%%%%%
\begin{figure}[th]
\begin{center}
\vspace*{-0.4cm}
\epsfig{figure=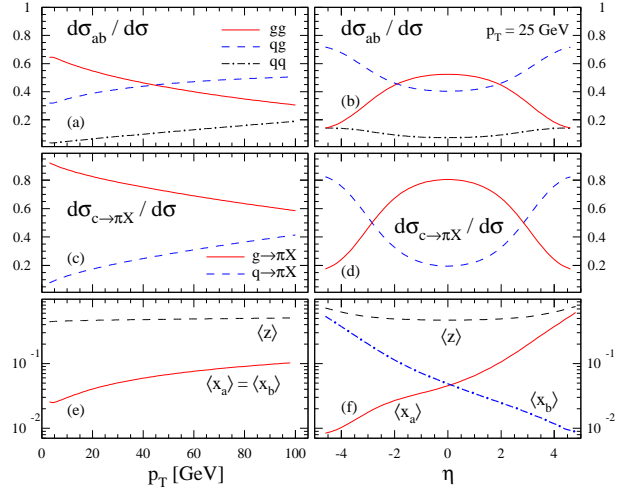,width=0.48\textwidth}
\end{center}
\vspace*{-0.5cm}
\caption{(a) and (b): relative contributions of NLO partonic subprocesses
$d\sigma_{ab}$ initiated by gluon-gluon, quark-gluon, and quark-quark
scattering to the $\pi^0$ cross section at $\sqrt{S}=7\,\mathrm{TeV}$ 
as a function of $p_T$ and $\eta$, respectively. (c) and (d): same as in (a) and (b)
but now for the relative fractions of quarks and gluons fragmenting
into the observed pion. (e) and (f): mean values of the momentum
fractions $x_{a,b}$ and $z$ probed in $\pi^0$ production.
\label{fig:pp-support}}
\end{figure}
These aspects are explained in some detail in Fig.~\ref{fig:pp-support}.
The panels show the relevance of the different
NLO partonic subprocesses $d\sigma_{ab}$ (upper row), the
relative fractions of quark and gluon fragmentation (middle row),
and the mean values of the momentum
fractions $x_{a,b}$ and $z$ (lower row) probed in $\pi^0$ production at
$\sqrt{S}=7\,\mathrm{TeV}$. The results on the left hand side (l.h.s.)
are given as
a function of $p_T$ for $\eta=0$, and the panels on right hand side (r.h.s.)
are differential in $\eta$ for fixed $p_T=25\,\mathrm{GeV}$.
For different c.m.s.\ energies, the plots roughly scale with 
$x_T=2p_T/\sqrt{S}$, i.e., subprocess fractions are similar for
the same value of $x_T$.

From panels (a) and (b) one can infer that gluon-gluon scattering
is the most important channel at relatively small $p_T$ and central
rapidities $\eta$. For $p_T\gtrsim 40\,\mathrm{GeV}$ at $\eta\simeq 0$
or at forward rapidities, quark-gluon scattering becomes the dominant
subprocess. For instance, at large forward rapidities,
$\eta\gg 0$, one is mainly sensitive to
the scattering of a valence quark carrying a large momentum fraction
$x_a$ off a gluon with $x_b\ll 1$ for which both PDFs are large; 
cf.\ also panel (f). 
In the shown kinematic range, quark-quark initiated processes are always
small and reach a level of $15\%$ only for large $p_T$ and/or large
pseudorapidities $\eta$.

Figures~\ref{fig:pp-support} (c) and (d) give the relative contributions of
quark and gluon fragmentation to the $p_T$ and $\eta$ differential
cross sections, respectively. At mid rapidity, gluon-to-pion fragmentation
is dominant in the entire range of $p_T$ shown. For $|\eta|\gg 0$ and 
$p_T=25\,\mathrm{GeV}$, quark fragmentation reaches a level of $80\%$.
We note that for smaller values of $p_T$, the share between quark and gluon
induced hadron production is roughly equal. 

In the lower panels of Fig.~\ref{fig:pp-support} we present estimates
of the mean momentum fractions $\langle x_{a,b}\rangle$ and
$\langle z\rangle$ which are predominantly probed in single-inclusive
hadron production at the LHC.
There are several ways to estimate, for instance, an average $\langle z\rangle$.
We define it in the standard way by evaluating the convolutions in
Eq.~(\ref{eq:xsec}) with an extra factor of $z$ in the integrand,
divided by the cross section itself \cite{ref:guzey}, i.e., schematically
we use
\begin{equation}
\label{eq:mean}
\langle z\rangle \equiv
\frac{\int dz \,z\, \frac{d\sigma^H}{dzdp_T}}
{\int dz \, \frac{d\sigma^H}{dzdp_T}}\,.
\end{equation}
%
%%%%%%%%%%%%%%%%%%%%%%%%%%%%%%%%%%%%%%%%%%%%%%%%%%%%%%%%%%%%%%%
% FIGURE 7: XT SCALING
%%%%%%%%%%%%%%%%%%%%%%%%%%%%%%%%%%%%%%%%%%%%%%%%%%%%%%%%%%%%%%%
\begin{figure}[th]
\begin{center}
\vspace*{-0.6cm}
\epsfig{figure=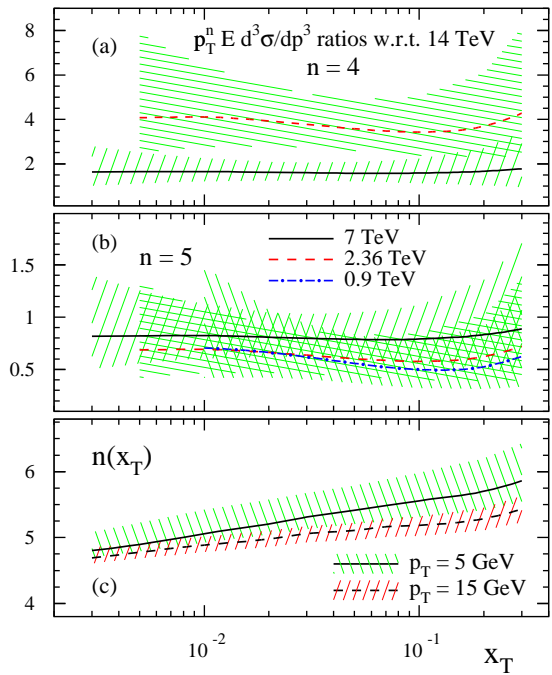,width=0.48\textwidth}
\end{center}
\vspace*{-0.5cm}
\caption{Illustration of $x_T$ scaling for $\pi^+$ production
at $\eta=0$ and LHC energies.
(a) and (b): ratios of the invariant cross section 
in Eq.~(\ref{eq:xtscaling}) scaled by $p_T^n$ for
$7\,\mathrm{TeV}$ (solid line), $2.36\,\mathrm{TeV}$ (dashed line), 
and $0.9\,\mathrm{TeV}$ (dot-dashed line) to the result obtained
for $\sqrt{S}=14\,\mathrm{TeV}$ for $n=4$ and $n=5$, respectively.
The bands indicate the theoretical uncertainty from varying the
scales $\mu_{f,{f^{\prime}},r}$ independently in the numerator and
denominator of the ratios. 
(c): $x_T$ dependence of the scaling exponent $n$ 
and scale uncertainty (shaded bands) estimated
based on Eq.~(\ref{eq:nexp}).
\label{fig:xt}}
\end{figure}
$\langle x_{a,b}\rangle$ are estimated accordingly.
The most important thing to notice is that despite the large c.m.s.\ energies
available at the LHC, the bulk of the hadrons is produced with $\langle z\rangle \ge 0.4$
and $\langle z\rangle \ge 0.6$ for $\eta=0$ and $\eta\gg 1$, respectively,
which is comparable to what one finds at much lower c.m.s.~energies, for instance,
at RHIC. This important finding ensures the applicability of the concept of FFs
which, as discussed in the Introduction, requires $z\gtrsim 0.05$.
Another interesting result is that at mid rapidity the average $x_{a,b}$
is considerably larger than what one might naively expect from 
the lower kinematic limit $x_{a,b}\simeq x_T=2p_T/\sqrt{S}$ which is
of the order a few times $10^{-3}$ for $p_T\lesssim 20\,\mathrm{GeV}$.

Figures \ref{fig:piplus}-\ref{fig:pp-support} clearly demonstrate the potential
impact of single-inclusive hadron measurements in a wide kinematic
range in $p_T$ and $\eta$ on future FF global analyses. 
Data will impose stringent constraints on the FFs for various hadron species
in the large $z$ range at unprecedented large scales $\mu_{f^{\prime}}\simeq p_T$.
Scale and to some extent also PDF uncertainties are considerably smaller than 
for corresponding measurements at RHIC performed at considerably smaller values of $p_T$
and $\sqrt{S}$ which are currently used in global fits \cite{ref:dsspion,ref:dssproton}.

Finally, we discuss the property of $x_T$ scaling at LHC energies.
In pQCD one expects deviations from the naive power-law scaling of the invariant
cross section $\sigma_{\mathrm{inv}}$ in Eq.~({\ref{eq:xtscaling}) leading to
$n=n(x_T,\sqrt{S})$ \cite{ref:xt,ref:arleo}.
Figures~\ref{fig:xt} (a) and (b) show ratios of $\sigma_{\mathrm{inv}}$
scaled by $p_T^n$ for $7\,\mathrm{TeV}$, $2.36\,\mathrm{TeV}$, 
and $0.9\,\mathrm{TeV}$ to the result obtained for $\sqrt{S}=14\,\mathrm{TeV}$ 
for fixed $n=4$ and $n=5$ in Eq.~(\ref{eq:xtscaling}), respectively.
As can be seen, scaling violations are sizable for $n=4$ as should be expected
from the running of $\alpha_s$ and the scale evolution of PDFs and FFs.
Note that the ratio for $0.9\,\mathrm{TeV}$ is too large to be displayed
in Fig.~\ref{fig:xt}~(a). The choice $n=5$ in Eq.~(\ref{eq:xtscaling}) leads
to much more similar ratios for all three energies in the entire range of $x_T$, 
in particular for $\sqrt{S}=2.36\,\mathrm{TeV}$ and $0.9\,\mathrm{TeV}$.

Power-law scaling of $\sigma_{\mathrm{inv}}$ with a universal exponent
$n$ can, however, never be more than a rough approximation because $n$ must
depend on both $p_T$ and $\sqrt{S}$ rather than being constant. 
This kinematic dependence of $n$ is illustrated in 
Fig.~\ref{fig:xt}~(c) for two different values of $p_T$.
As in Ref.~\cite{ref:arleo}, we estimate the scaling exponent $n$ in Eq.~(\ref{eq:xtscaling})
by comparing $x_T$ spectra at different c.m.s.~energies $\sqrt{S}$ and $\sqrt{S^\prime}$
for fixed $p_T$, i.e.,
\begin{equation}
n(x_T)=-
\frac{\ln \left[ \sigma_{\mathrm{inv}}(S,x_T)/\sigma_{\mathrm{inv}}(S^{\prime},x_T) \right]}
{\ln(\sqrt{S}/\sqrt{S^{\prime}})}\,.
\label{eq:nexp}
\end{equation}
A dependence on both $x_T$ and $p_T$ is clearly visible and needs to be taken into account
when comparing theoretical expectations for $x_T$ scaling to experimental spectra.
Consequently, there is also some ambiguity in estimating $n(x_T)$ as the result 
based on Eq.~(\ref{eq:nexp}) depends also on the choice of $S$ and $S^{\prime}$.

In addition, studies of the scaling behavior also suffer, of course, from theoretical scale
ambiguities. The shaded bands in Figs.~\ref{fig:xt}~(a) and (b) indicate
the uncertainty from varying the scales $\mu_{f,{f^{\prime}},r}$ by the usual 
factor of two up and down the default choice $p_T$.
Compared to Ref.~\cite{ref:arleo}, we find a much larger variation
because, as a more conservative choice, we allow for {\em different}
scales in the numerator and denominator of the ratios of invariant
cross sections shown in panels (a) and (b).

The possibility of having different factorization scales for observables
calculated at different c.m.s.\ energies is, e.g., natural in certain 
proposed ``scale fixing" procedures \cite{ref:blm}, as the ``optimum scale" 
depends on the kinematics of the process.
It is also interesting to recall recent measurements of single-inclusive
neutral pion production at RHIC for three different c.m.s.~energies \cite{ref:rhicpion}.
Although the cross sections are described well by pQCD within the sizable
scale uncertainties, it can be argued that an ``optimized" choice of scales 
in a NLO calculation \cite{ref:xsecnlo} would suggest to use scales closer to
$p_T/2$, $p_T$, and $2p_T$ for $\sqrt{S}=62.4$, $200$, and $500\,\mathrm{GeV}$,
respectively. The smaller factorization scale at lower c.m.s.~energies
allows for more QCD radiation in the hard scattering matrix elements 
which increases the cross section and leads to a more favorable description of the data.
To some extent this mimics all order resummations of logarithmic
contributions in the partonic subprocess cross sections
which are enhanced near the partonic threshold and hence more relevant at lower
c.m.s.~energies for a given value of $p_T$ \cite{ref:resum}.

%%%%%%%%%%%%%%%%%%%%%%%%%%%%%%%%%%%%%%%%%%%%%%%%%%%%%%%%%%%%%%%%
\subsection{Unidentified charged hadron spectra}
%%%%%%%%%%%%%%%%%%%%%%%%%%%%%%%%%%%%%%%%%%%%%%%%%%%%%%%%%%%%%%%%%
%
%%%%%%%%%%%%%%%%%%
% FIGURE 8: UNIDENTIFIED HADRONS
%%%%%%%%%%%%%%%%%%
\begin{figure}[th]
\begin{center}
\vspace*{-0.6cm}
\epsfig{figure=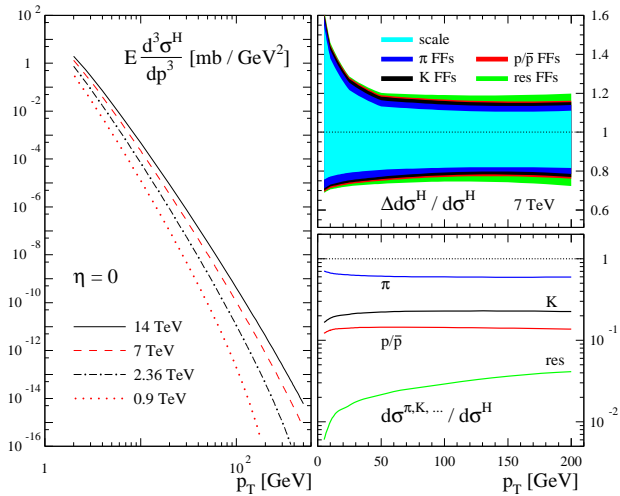,width=0.48\textwidth}
\end{center}
\vspace*{-0.5cm}
\caption{Left panel: NLO invariant cross section for the production of charged hadrons 
for $\eta=0$ at four different c.m.s.\ energies. Upper right panel: relative
theoretical ambiguity of the charged hadron yield at $\sqrt{S}=7\,\mathrm{TeV}$ 
induced by variations of the scales $\mu_{f,{f^{\prime}},r}$ and the 
uncertainties of the FFs for the individual hadron species. 
Lower right panel: Partial contribution of the different hadronic species to
the the invariant charged hadron cross section at $\sqrt{S}=7\,\mathrm{TeV}$.
\label{fig:lhc-hadron}}
\end{figure}
In the following, we provide a detailed assessment of how the present 
limitations of our knowledge of FFs 
propagate to uncertainties for cross section estimates.
We focus the discussions on unidentified charged hadron spectra, 
which, in addition to be accessible with all LHC detectors as well as
the Tevatron experiments, allow us to compare the relative contributions 
of different hadron species and their respective uncertainties. 

Pions clearly dominate the charged hadron spectra, with
typical uncertainties for the relevant FFs $D_c^{\pi^{\pm}}$
estimated to be at the few percent level \cite{ref:dsspion}.
Nevertheless, charged kaons, protons, and anti-protons represent non-negligible contributions
and, potentiated by the much larger uncertainties inherent to
their FFs \cite{ref:dsspion,ref:dssproton}, may yield a significant contribution to the theoretical error for
estimates of charged particle spectra, perhaps even comparable in size 
to the one for pions.
Contributions from ``residual charged hadrons", i.e., hadrons other than
pions, kaons, and protons, are known to be completely 
marginal at low energy scales but increase with energy.
Since the corresponding FFs are very poorly known \cite{ref:dssproton}, 
they can produce sizable uncertainties, comparable or even larger than those coming 
from the much more copiously produced hadron species. 
Altogether, the combined theoretical error from FFs may become 
comparable to scale uncertainties in certain kinematic regions, in particular,
at larger transverse momenta.

To a first approximation, uncertainties from FFs are often estimated
by simply comparing the results obtained with two different optimum fits of FFs.
Such a procedure can give, however, at best a lower bound on the true
error. In order to arrive at a faithful estimate of uncertainties derived from
those inherent to FFs, we use the robust Lagrange multiplier technique
\cite{ref:lagrange}, which explores for any desired observable 
depending on FFs its full range of variations within chosen the range of
$\Delta \chi^2$ tolerated in the fit. 
Although the actual error estimate is more involved than in the standard
Hessian method \cite{ref:hessian}, it has the advantage of not
making any assumptions on the shape of the $\chi^2$ profile near the minimum of the fit
or on how the errors of the fit parameters describing the FFs propagate to
a given observable.
%
%%%%%%%%%%%%%%%%%%
% FIGURE 9: CDF
%%%%%%%%%%%%%%%%%%
\begin{figure}[th]
\begin{center}
\vspace*{-0.6cm}
\epsfig{figure=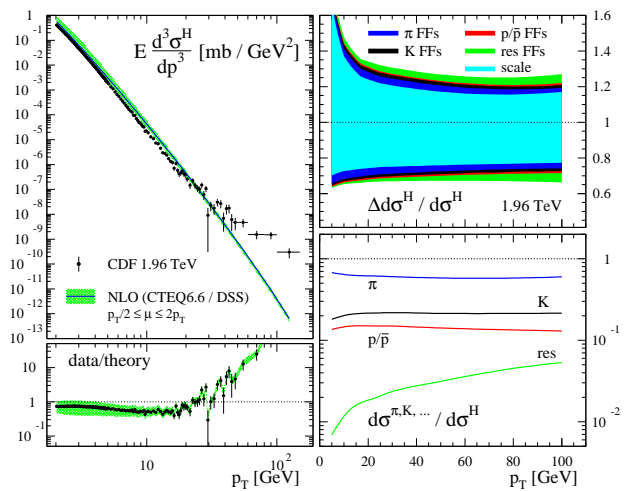,width=0.48\textwidth}
\end{center}
\vspace*{-0.5cm}
\caption{The same as Fig.~\ref{fig:lhc-hadron} but 
now for $p\bar{p}$ collisions at $1.96\,\mathrm{TeV}$.
The left panels show also recent data from CDF \cite{ref:cdfdata} and 
the ratio of data and a NLO calculation.
\label{fig:cdf}}
\end{figure}

As an example, the upper right panels of Figs.~\ref{fig:lhc-hadron} and \ref{fig:cdf} 
show the outcome of propagating the uncertainties of 
the FFs for different hadron species with the Lagrange multiplier method to
the invariant charged hadron cross sections (given in the left panels) for both the 
LHC and the Tevatron at c.m.s.\ energies of 7
and $1.96\,\mathrm{TeV}$, respectively. 
As in the DSS analysis \cite{ref:dsspion,ref:dssproton}, we allow for a tolerance
of $\Delta \chi^2/\chi^2 = 2\%$ in each error analysis.  
The uncertainties resulting from FFs are compared to the theoretical ambiguity from
variations of the scales within the typical range
$p_T/2\le \mu_{f,{f^{\prime}},r} \le 2p_T$.
Although the scale uncertainties are dominant, 
errors propagated from FFs to the invariant charged hadron cross sections
are quite sizable, and their role becomes increasingly significant 
at larger values of $p_T$. This is similar to what was found in Fig.~\ref{fig:plus}
for PDF uncertainties.

In the lower right panels of Figs.~\ref{fig:lhc-hadron} and \ref{fig:cdf} we show the 
relative contributions of the different hadronic species to the inclusive
charged hadron spectrum as a function of $p_T$. 
It is worth noticing that the mixture of hadrons is fairly independent of $p_T$,
which is linked to constant average momentum fraction $\langle z\rangle$ 
observed in Fig.~\ref{fig:pp-support} (e). However, the
contribution from residual charged hadrons becomes increasingly relevant at larger values
of $p_T$ but still remains fairly small as compared to pions, kaons, and protons.

In Fig.~\ref{fig:cdf} we compare the estimates for invariant cross section at NLO 
accuracy to recent data from CDF \cite{ref:cdfdata}. In the lower left panel we also
show the ratio of data and theory. As can be seen, the theoretical results significantly
undershoot the data for $p_T\gtrsim 25\,\mathrm{GeV}$ which caused quite a stir
\cite{ref:cdffuzz}. Since the data are also in excess of the corresponding jet
measurement, the most likely explanation for the observed discrepancy is
an experimental problem. 
Nevertheless, it is interesting to know that the FF uncertainties estimated above
are by far too small to account for the hadron yield observed by CDF.

Finally, it is worth mentioning that when going from Tevatron to LHC kinematics there
is quite some reduction in the relative importance of the residual
factorization and renormalization scale dependence as compared to FF uncertainties.
This again illustrates the impact of upcoming LHC data on future global analyses.

%%%%%%%%%%%%%%%%%%%%%%%%%%%%%%%%%%%%%%%%%%%%%%%%%%%%%%%%%%%%%%%%%%%%%%%%%%
\section{Pion production in $pPb$ Collisions}
%%%%%%%%%%%%%%%%%%%%%%%%%%%%%%%%%%%%%%%%%%%%%%%%%%%%%%%%%%%%%%%%%%%%%%%%%%
%
In this Section we propose a set of hadron production measurements in 
$pPb$ collisions at the LHC as a tool to characterize and quantify nuclear modifications in 
the fragmentation process. The LHC is capable of providing $pPb$ collisions at a maximum
c.m.s.\ energy of about $8.8\,\mathrm{TeV}$ \cite{ref:lhcppb} although such a program is not envisioned 
in the initial phase. However, measurements in $dAu$ collisions 
at RHIC were instrumental in interpreting results obtained in heavy ion collisions, 
which are considerably more complicated to understand theoretically.

We note that leading order estimates of neutral pion production in 
$pPb$ collisions at the LHC were recently presented in Ref.~\cite{ref:wiedemann} but based on the
assumption that all nuclear modifications can be entirely absorbed into
nPDFs. In view of the known sizable medium induced effects on hadron production 
yields in, e.g., lepton-nucleus collisions \cite{ref:hermes}, which cannot be explained by nPDFs \cite{ref:nff},
such an approach is questionable and may not be adequate. 
In fact, in a recent paper \cite{ref:nff} the concept of medium modified fragmentation functions
(nFFs) was introduced within the standard factorized framework of pQCD. 
These novel nFFs, which obey ordinary timelike scale evolution \cite{ref:evol} and 
have been extracted in a global QCD analysis for pions and kaons, 
allow one to treat hard reactions with identified hadrons consistently at NLO accuracy
when combined with nPDFs. This is the approach we pursue here to compute predictions
for $pPb$ collisions at the LHC.

Conventional factorization of short and long distance physics effects is, however, not expected
to hold in general in a nuclear environment \cite{ref:nuclreview,ref:breaking}, and indeed  
collisions between two heavy nuclei show a very large breaking pattern. However, 
factorization was shown to be phenomenologically very successful in describing current 
lepton-nuclei and deuteron-nuclei 
collision data, including their $A$ dependence \cite{ref:nff}. This supports the concept of 
universal, medium modified PDF and FFs at least at an approximate level. $pPb$ collisions 
at the LHC will certainly explore the limits of characterizing nuclear modifications
in a factorized pQCD approach.
%
%%%%%%%%%%%%%%%%%%%%%%%%%%%%%
% FIGURE XX: NPDF/NFF RATIOS
%%%%%%%%%%%%%%%%%%%%%%%%%%%%%
\begin{figure}[th]
\begin{center}
\vspace*{-0.6cm}
\epsfig{figure=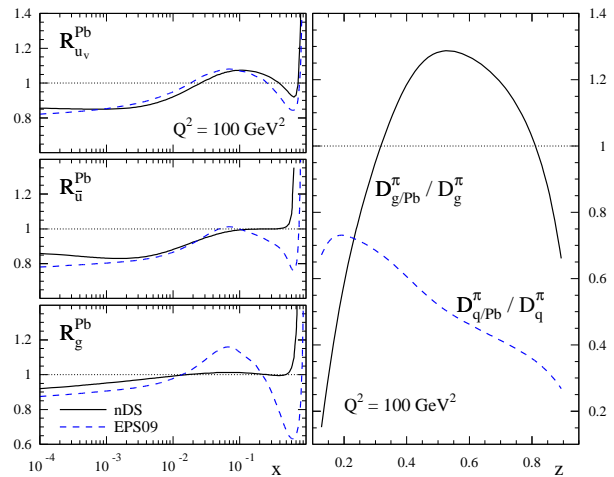,width=0.48\textwidth}
\end{center}
\vspace*{-0.5cm}
\caption{Nuclear modification factors in lead for PDFs \cite{ref:nds,ref:eps} (left panels) 
and FFs \cite{ref:nff} (right panel)
of different flavors at $Q^2=100\,\mathrm{GeV}^2$ 
as determined in recent global QCD analyses at NLO accuracy.
\label{fig:npdfnff}}
\end{figure}

To recall the pattern and magnitude of observed medium induced modifications 
for quarks and gluons,
we show in Fig.~\ref{fig:npdfnff} the ratios of nPDFs and nFFs
\begin{equation}
R^A_i(x,Q^2) \equiv \frac{f_i^A(x,Q^2)}{f_i^p(x,Q^2)},\;
R^H_{i/A}(z,Q^2) \equiv \frac{D^H_{i/A}(z,Q^2)}{D_i^H(z,Q^2)}
\end{equation}
to the standard PDFs and vacuum pion FFs, respectively.
The ratios are evaluated for $Pb$ at a scale of $Q^2=100\,\mathrm{GeV^2}$ relevant
for pion production at transverse momentum $p_T\simeq 10\mathrm{GeV}$.

The l.h.s.~of Fig.~\ref{fig:npdfnff} shows the nuclear modification factors
$R^{Pb}_{i}$ for $u$ valence quarks, $\bar{u}$ sea quarks, and
the gluon $g$ for two standard NLO sets of nPDFs: nDS \cite{ref:nds} (solid lines) 
and EPS09 \cite{ref:eps} (dashed lines).
The deviations of $R^{Pb}_{i}$ from unity depend on the $x$ region 
and are typically referred to as ``shadowing" ($x\lesssim 0.01$), 
``anti-shadowing" ($0.01\lesssim x\lesssim 0.2$), ``EMC effect" ($0.2\lesssim x\lesssim 0.7$),
and ``Fermi motion" ($x\gtrsim 0.7$) in the terminology of nuclear DIS.
Differences between the EPS09 and nDS sets are most pronounced for the gluon nPDF, which
plays a major role in $pPb$ collisions. Contrary to PDFs, the behavior of nPDFs
is basically unconstrained below $x\simeq 0.01$ by present data, an $x$ region 
particularly prone to possible novel, non-linear features of QCD scale evolution.

The corresponding medium modifications for FFs are displayed in the right panel of
Fig.~\ref{fig:npdfnff} and are distinctly different for quarks and gluons, where one finds 
suppression and enhancement, respectively, compared to vacuum fragmentation functions.
The pattern is readily explained by the dominant role of quark fragmentation in
describing the observed hadron attenuation in SIDIS off a heavy nucleus, while
the enhancement of hadrons in $dAu$ collisions is closely linked with the gluon nFF \cite{ref:nff}.
In general, the observed nuclear effects are more pronounced for nFFs than for nPDFs, 
and nFFs can either enhance or overturn medium modifications computed with nPDFs 
but vacuum FF as was done, e.g., in Ref.~\cite{ref:wiedemann}. 

%%%%%%%%%%%%%%%%%%%%%
% FIGURE 11: pPb xsec
%%%%%%%%%%%%%%%%%%%%%
\begin{figure}[th]
\begin{center}
\vspace*{-0.6cm}
\epsfig{figure=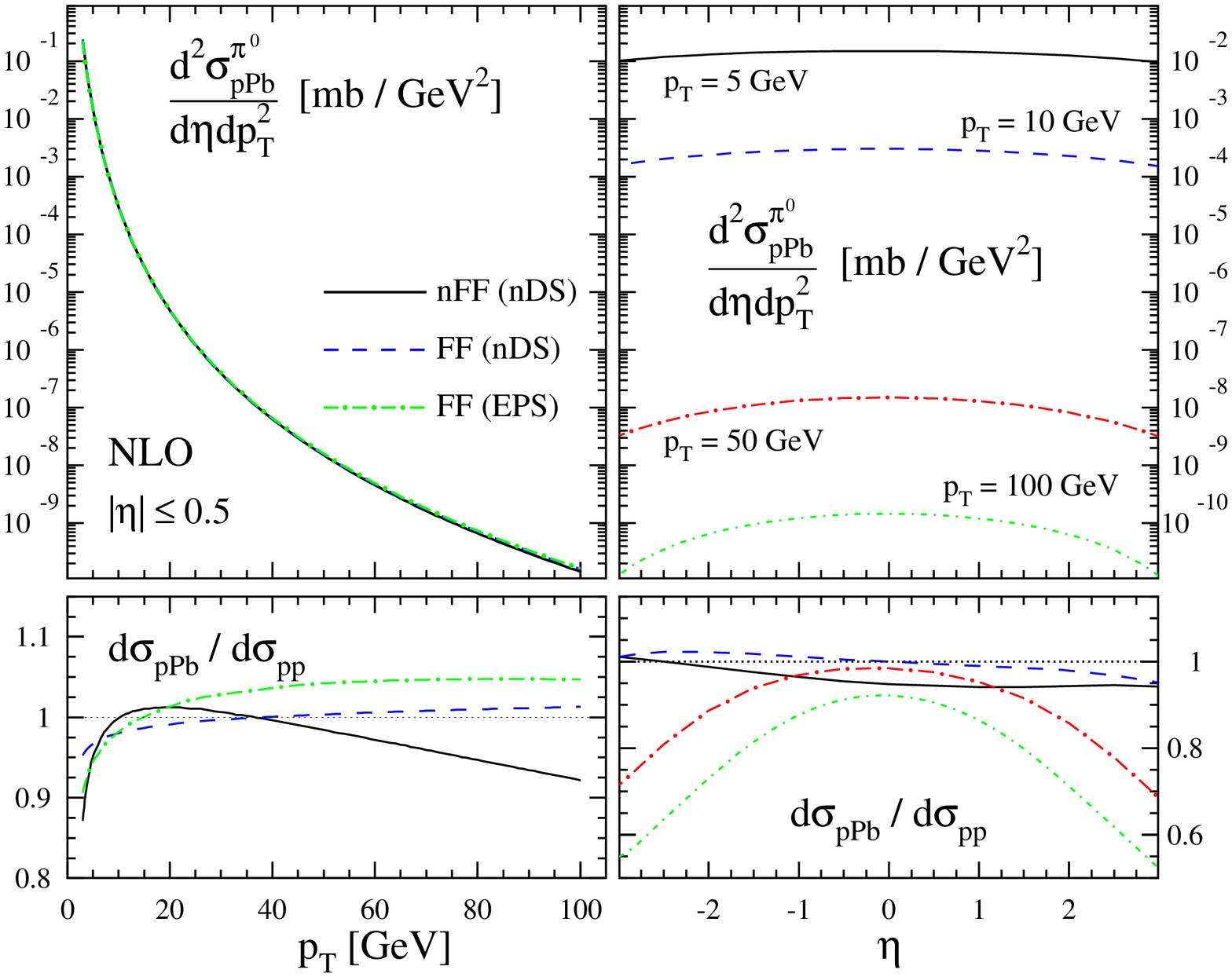,width=0.48\textwidth}
\end{center}
\vspace*{-0.5cm}
\caption{Transverse momentum (left panels) and pseudorapidity (right panels) dependence of neutral pion production in $pPb$ collisions at $\sqrt{S}=8.8\,\mathrm{TeV}$.
The upper panels show the cross sections at NLO accuracy, 
and lower panels display the ratio to the corresponding
cross section in $pp$ collisions.
The $p_T$ dependent yields are integrated in $|\eta|\le 0.5$ and
computed for different combinations of nPDFs \cite{ref:nds,ref:eps}, 
nFFs \cite{ref:nff}, and vacuum FFs \cite{ref:dsspion}.
The $\eta$ differential results are given for various fixed values of $p_T$, and
were obtained using the nDS set of nPDFs \cite{ref:nds} and nFFs of \cite{ref:nff}.
\label{fig:pbxsec}}
\end{figure}
Assuming that factorization holds in $pPb$ collisions at a c.m.s.\ energy of
$\sqrt{S}=8.8\,\mathrm{TeV}$, we present predictions for the production
of neutral pions in Fig.~\ref{fig:pbxsec} at NLO accuracy using nPDFs and
the recently proposed set of nFFs \cite{ref:nff} in Eq.~(\ref{eq:xsec}).
By construction, the entire nuclear dependence resides in the non-perturbative
nPDFs and nFFs, and both the scale evolution and the partonic
hard scattering cross sections $d\hat{\sigma}_{ab\to c X}$
are taken to be the same as in $pp$ collisions.

The upper left panel shows the $p_T$ dependence at central pseudorapidities
$|\eta|\le 0.5$ using $\mu_f=\mu_{f^{\prime}}=\mu_r=p_T$ in Eq.~(\ref{eq:xsec}). 
We refrain from showing alternative results for different choices 
of $\mu_{f,f^{\prime},r}$ as the theoretical scale ambiguities are very similar to the
ones for $pp$ collisions discussed in Sec.~\ref{sec:pp}.
Because of the steep fall of the cross section with $p_T$ over several orders
of magnitude any differences between the NLO calculations based
on various combinations of nPDFs and standard vacuum or medium modified FFs are hard to notice. 
The different patterns of nuclear modification become clearly visible, however, 
when expressed as ratios to the corresponding cross section in $pp$ collisions 
at the same c.m.s.\ energy, which are given 
in the lower left panel of Fig.~\ref{fig:pbxsec}.

The impact of medium induced effects on the hadronization process is clearly visible
at larger $p_T$ by comparing the results obtained with the nDS set of nPDFs \cite{ref:nds}
along with nFFs \cite{ref:nff} (solid line) and with DSS vacuum FFs \cite{ref:dsspion} 
(dashed line), leading to hadron attenuation and enhancement, respectively.
The uncertainties in our present knowledge of nPDFs can be inferred from 
comparing the dot-dashed curve, which uses again the DSS vacuum FFs but 
now the EPS09 set of nPDFs \cite{ref:eps}, with the dashed line. The EPS nPDFs result in a somewhat
more sizable enhancement than the nDS set due to the larger $R_g^{Pb}$ at medium $x$ values,
see Fig.~\ref{fig:npdfnff}, which are predominantly probed here; cf.\ Fig.~\ref{fig:meanxz} below.
Notice that the EPS09 analysis \cite{ref:eps} includes $dAu\to \pi^0X$ data from RHIC
by assuming, however, that the entire nuclear dependence resides 
only in the initial state, i.e., in the nPDFs. This is clearly inadequate in view
of the sizable hadron attenuation observed in semi-inclusive lepton-nucleus scattering \cite{ref:hermes}.
Hence, in what follows we will only use the NLO nDS set of nPDFs, which was also obtained in
a convolutional approach, very similar to the nFF analysis \cite{ref:nff}
based on Eq.~(\ref{eq:conv-ansatz}).
  
The panels on the r.h.s.~of Fig.~\ref{fig:pbxsec} show
the pseudorapidity dependence of the $\pi^0$ yields at NLO accuracy
for various fixed values of $p_T$. All results are obtained with the nDS set
of nPDFs and nFFs of \cite{ref:nff}. As before, we choose 
$\mu_f=\mu_{f^{\prime}}=\mu_r=p_T$ in Eq.~(\ref{eq:xsec}).
The upper panel displays the differential cross sections, and the lower one the 
ratios to the corresponding results in $pp$ collisions.
The rapidity dependence of the $pPb$ nuclear modification factor is of great phenomenological
relevance as it probes different ranges of momentum fraction $x_b$ in the nPDFs while
the average value of $z$ in the fragmentation process only 
slowly increases with larger $|\eta|$; cf.~Fig.~\ref{fig:pp-support}.
In addition, larger values of $p_T$ at fixed $\eta$ probe both on average larger $x_{a,b}$ and $z$ values.
Pions produced in the backward ($\eta<0$) direction of the proton beam require large
momentum fractions $x_b$, and nPDFs are mainly probed in the anti-shadowing and EMC region.
In contrast, forward hadrons are sensitive to smaller values of $x_b$ where shadowing is
expected; see Fig.~\ref{fig:npdfnff}. In general, this leads to nuclear modification
factors which are asymmetric in rapidity for any given value of $p_T$.

Based on these kinematic considerations, it was argued in
Ref.~\cite{ref:wiedemann} that a detailed study of the $p_T$ and $\eta$ dependent hadron
yields at the LHC will help to determine nPDFs more precisely. 
However, the neglected medium modifications on the FFs can change the
theoretical expectations for the ratio $d\sigma_{pPb}/d\sigma_{pp}$ 
significantly, as can be seen in Fig.~\ref{fig:pbxsec}.
In particular, at larger $p_T$ and rapidities our expectations
based on consistently including also nFFs in Eq.~(\ref{eq:xsec})
show much larger nuclear modification factors than the few percent
effects estimated in Fig.~2 of \cite{ref:wiedemann}. This is readily explained by the
strong depletion of the quark FFs at large values of $z$
as displayed in Fig.~\ref{fig:npdfnff}, which is most
relevant at large $\eta$ and $p_T$.

%%%%%%%%%%%%%%%%%%
% FIGURE 12: NFF SUPPORT PLOT
%%%%%%%%%%%%%%%%%%
\begin{figure}[th]
\begin{center}
\vspace*{-0.6cm}
\epsfig{figure=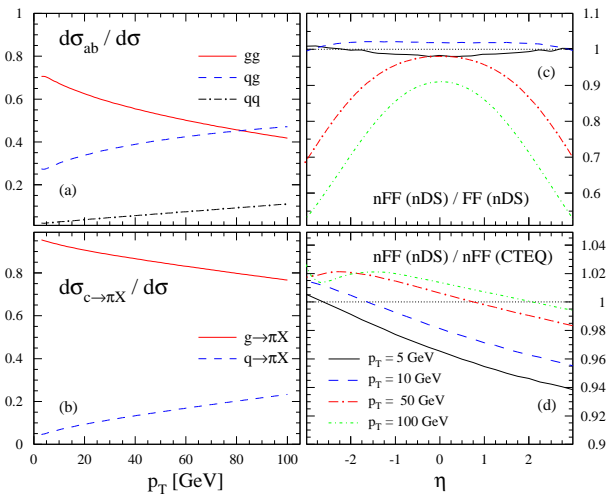,width=0.48\textwidth}
\end{center}
\vspace*{-0.5cm}
\caption{(a): relative contributions of NLO partonic subprocesses $d\sigma_{ab}$ initiated by  
gluon-gluon, quark-gluon, and quark-quark scattering to the $pPb$ cross section
shown in Fig.~\ref{fig:pbxsec}. (b): same as in (a) but now for the relative fractions
of quarks and gluons fragmenting into the observed pion.
(c) and (d) impact of medium modified FFs and nPDFs, respectively,
on the rapidity dependence of the nuclear modification factors 
for different values of $p_T$ shown in the  lower right
 panel of Fig.~\ref{fig:pbxsec}.
\label{fig:nffsupport}}
\end{figure}
To elucidate this further, Fig.~\ref{fig:nffsupport} gives more details 
on the relevance of different partonic subprocesses (a),
the role of quarks and gluons in the hadronization (b), and the impact of
nFFs (c) and nPDFs (d) on the rapidity dependence of the nuclear modification 
factor $d\sigma_{pPb}/d\sigma_{pp}$. 

Panel (a) illustrates that in the entire $p_T$ range shown in Fig.~\ref{fig:pbxsec}
gluon initiated subprocesses dominate the cross section at 
central rapidities. Most of the pions are produced from
gluon fragmentation as can been seen in Fig.~\ref{fig:nffsupport}~(b).
Only for $p_T\gtrsim 50\,\mathrm{GeV}$ quark nFFs contribute
at the level of $20\%$ or more.
For hadrons produced at larger $\eta$, the fractions in (a) and (b) shift
somewhat in favor of quark-gluon scattering and quark fragmentation, respectively.

Figures \ref{fig:nffsupport} (c) and (d) make explicit that the dominant
medium effect on $d\sigma_{pPb}/d\sigma_{pp}$ 
resides indeed in the final-state. Panel (c) shows the ratio where the denominator
is computed with the nDS set of nPDFs such that the entire deviation from
unity is due to the nFFs. Likewise, in panel (d) we compute $d\sigma_{pp}$ 
in the nuclear modification factor with nFFs rather than vacuum FFs. This 
quantifies medium effects due to the initial-state nPDFs. The latter effect
is much smaller, and our results agree with the LO estimates shown in Ref.~\cite{ref:wiedemann}.
Thus we believe that single inclusive hadron production at the LHC 
will provide a decisive test of the proposed
factorized framework for $pPb$ collisions and the concept of medium modified
FFs. In particular, the ALICE experiment is capable of identifying different
hadron species in a wide range of rapidity which will greatly facilitate 
such theoretical studies.
To disentangle and characterize nuclear effects in PDFs and in FFs
precisely, requires to study also other hard processes in $pPb$ collisions
not affected by hadronization like jet production, Drell-Yan, or prompt photons. 
Such measurements will provide an important input to future global QCD analyses
of nPDFs.

%%%%%%%%%%%%%%%%%%
% FIGURE 13: NFF SUPPORT PLOT: MEAN X,Z
%%%%%%%%%%%%%%%%%%
\begin{figure}[th]
\begin{center}
\vspace*{-0.8cm}
\epsfig{figure=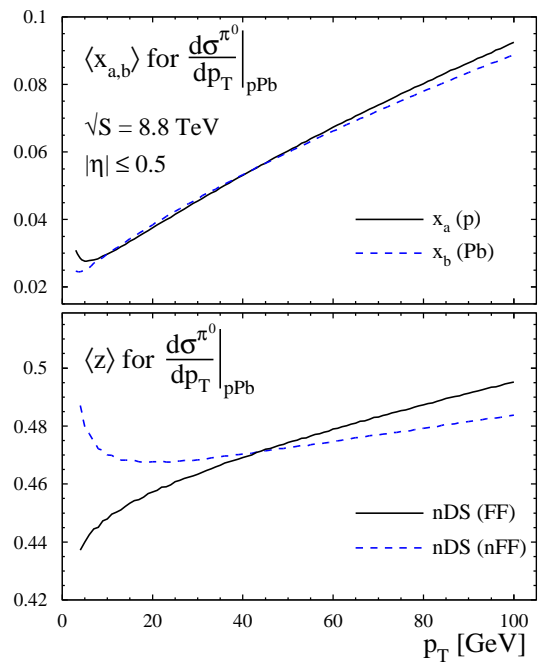,width=0.46\textwidth}
\end{center}
\vspace*{-0.5cm}
\caption{Upper panel: mean values of $x_a$ and $x_b$ probed in 
the proton and the lead nucleus, respectively, 
for pion production at central rapidities in
$pPb$ collisions at $\sqrt{S}=8.8\,\mathrm{TeV}$. 
Lower panel: same as above but now for 
the mean value of $z$ computed with the nDS set of nuclear PDFs
and DSS vacuum FFs \cite{ref:dsspion}
(solid line) and nFFS of Ref.~\cite{ref:nff} (dashed line).
\label{fig:meanxz}}
\end{figure}

Finally, Fig.~\ref{fig:meanxz} gives an idea which values of
$x_{a,b}$ and $z$ are probed on average in the $p_T$ differential
cross section at central rapidities shown in Fig.~\ref{fig:pbxsec}.
As for $pp$ collisions, the estimates are based on Eq.~(\ref{eq:mean}).
It turns out that
measurements of pion production up to transverse momenta of
$100\,\mathrm{GeV}$ probe both the nPDFs and the usual PDFs in
the range of momentum fractions from 0.01 to 0.1. To access smaller
values in the lead nucleus, where novel, non-linear features of
QCD scale evolution may become relevant, 
one needs to go to very forward pseudorapidities \cite{ref:guzey}.
From the lower panel of Fig.~\ref{fig:meanxz} one can infer that
$pPb$ collisions are mainly sensitive to fairly large values of
momentum fraction taken by the produced pion, with $\langle z\rangle$ slightly
increasing with $p_T$. This is similar to what was observed 
in $dAu$ collisions at RHIC energies \cite{ref:nff} and for $pp$ collisions
above in Fig.~\ref{fig:pp-support}.
Despite the large c.m.s.\ energy of $\sqrt{S}=8.8\,\mathrm{TeV}$, 
(n)FFs can be safely applied as one
is fairly insensitive to the region of small $z$, where the
concept of (n)FFs is bound to fail due to finite hadron mass effects,
higher twist contributions, and the singular behavior of the timelike
evolution kernels. We note that one samples slightly different
values $\langle z\rangle$ in the nuclear and vacuum FFs as can been
seen by comparing the dashed and solid lines.

%%%%%%%%%%%%%%%%%%%%%%%%%%%%%%%%%%%%%%%%%%%%%%%%%%%
\section{\label{sec:conclu} Summary and Conclusions}
%%%%%%%%%%%%%%%%%%%%%%%%%%%%%%%%%%%%%%%%%%%%%%%%%%%%
%
We have presented a comprehensive analysis of single-inclusive hadron
production in $pp$ and $pPb$ collisions at LHC energies based on QCD factorization.

It was shown that first results from the LHC experiments for charged hadron spectra 
agree well with expectations based on NLO pQCD calculations using latest sets of 
parton distribution and fragmentation functions. 

Based on this success, we have given detailed
predictions for various kinematic distributions of identified and unidentified
hadrons in $pp$ collisions. Different sources of theoretical uncertainties were discussed and 
estimated. It turned out that the residual factorization and renormalization dependence of the
cross sections at NLO accuracy represents the dominant source of uncertainty, in particular, 
at small values of transverse momentum of the produced hadron.
Approximate $x_T$-scaling with $n\simeq 5$ can be accommodated within
the rather large theoretical scale ambiguity.
Uncertainties from fragmentation functions for different hadron species 
were propagated to the single-inclusive yields based on the robust Lagrange 
multiplier method and found to be sizable, although smaller than those 
associated with the truncation of the perturbative series at NLO. 

To elucidate the possible impact of upcoming hadron production data from the LHC 
on future global analyses of fragmentation functions, we have studied in detail the
relative contributions of different partonic subprocesses and the fractions
of quark and gluon fragmentation into the observed hadrons as functions of 
transverse momentum and pseudorapidity. In addition, we have estimated the mean
values of momentum fractions both in the parton densities functions 
and in the fragmentation process which are relevant at LHC energies. 
It was found that like at hadron colliders at lower energies, most of the produced
hadrons take a rather large fraction of the parent parton's momentum which
ensures the applicability of the concept of factorized fragmentation functions
also at LHC energies.

Finally, we have proposed a set of measurements of single-inclusive hadron production
in proton-lead collisions to shed light on the so far poorly understood 
hadronization mechanism in a nuclear medium. 
We have presented expectations for pion yields at NLO accuracy
based on standard QCD factorization using
sets of medium modified parton distribution and fragmentation functions.

Single-inclusive hadron production in an unprecedented energy range at the LHC will
challenge our current understanding of fragmentation functions and help to further
constrain them, in particular, at large momentum fractions. 
The large range of transverse momenta of the produced hadrons 
will allow for detailed studies of the scale evolution for fragmentation functions.
Data obtained in proton-lead
collisions will explore the limits of characterizing nuclear modifications of
hadron production yields in a factorized QCD approach and scrutinize the 
applicability of the recently proposed concept of medium modified fragmentation functions.

%%%%%%%%%%%%%%%%
\section*{Acknowledgments}
%%%%%%%%%%%%%%%%
%
We thank Z. Trocsanyi and W. Bell for their help with the CMS and ATLAS 
data. This work was partially supported by CONICET, ANPCyT, UBACyT, BMBF, and 
the Helmholtz Foundation.

%%%%%%%%%%%%%%%%%%%%%%%%%%%

\end{document}